\documentclass[aps,prl,twocolumn,superscriptaddress,10pt]{revtex4-2}
\usepackage{bm}                 
\usepackage{graphicx}
\usepackage{xcolor}
\usepackage{amsmath}
\usepackage[T1]{fontenc}
\usepackage{textcomp}
\usepackage{amsfonts,amsmath,amscd,amssymb}
\usepackage{esint}
\usepackage{natbib}
\usepackage{wasysym}
\usepackage[normalem]{ulem}
\usepackage[version=3]{mhchem} 
\usepackage{braket}
\usepackage{tikz}
\usetikzlibrary{shapes,positioning}
\usepackage{CJK}
\newcommand{\sgn}{\text{sgn}}
\newcommand{\rev}[1]{\textcolor{black}{#1}}
\newcommand{\revv}[1]{\textcolor{black}{#1}}
\newcommand{\Chula}{Department of Physics, Faculty of Science, Chulalongkorn University, Patumwan, Bangkok 10330, Thailand}

\DeclareMathAlphabet\mathbfcal{OMS}{cmsy}{b}{n}

\makeatletter
\let\selectlanguage\@gobble
\makeatother

\begin{document}

\title{Probing the Potential Profile of Twisted Bilayer Graphene via Fabry-Pérot Interference}

\author{Curtis McDowell}
\thanks{These authors contributed equally.}
\affiliation{\Chula}

\author{Alina Mreńca-Kolasińska}
\thanks{These authors contributed equally.}
\affiliation{Faculty of Physics and Applied Computer Science, AGH University of Krakow, al. Mickiewicza 30, Krak\'ow 30-059, Poland}

\author{Kenji Watanabe}
\affiliation{Research Center for Electronic and Optical Materials, National Institute for Materials Science, 1-1 Namiki, Tsukuba 305-0044, Japan}

\author{Takashi Taniguchi}
\affiliation{Research Center for Materials Nanoarchitectonics, National Institute for Materials Science,  1-1 Namiki, Tsukuba 305-0044, Japan}

\author{Ming-Hao Liu}
\affiliation{Center for Quantum Frontiers of Research and Technology (QFort), National Cheng Kung University, Tainan 70101, Taiwan}

\author{Thiti Taychatanapat}
\email[]{thiti.t@chula.ac.th}
\affiliation{\Chula}

\date{\today}
\begin{abstract}
	  We use Fabry-Pérot interference to probe the internal potential profile of large-angle twisted bilayer graphene. We trace anomalous resistance oscillations in the nominally unipolar regime to a hidden cavity formed by unintentional local doping in our device. By analyzing zero-field interference patterns, we determine the location and size of this inhomogeneity. Magnetotransport measurements support the model, distinguishing  local cavity modes from global resonances through their different magnetic dependence. Simulations using our extracted profile reproduce the experimental features. Our results highlight interference spectroscopy as a simple, non-invasive probe for identifying local defects and internal potential barriers in ballistic devices.
\end{abstract}

\maketitle

\section{Introduction}

Fabry-Pérot (FP) interference is a phase-coherent phenomenon that applies to not only optical but also electron waves, and has been observed in clean electronic devices, including gate-defined graphene p-n junctions~\cite{Young_2009,Campos2012,Grushina2013,Rickhaus2013}. It has been used to investigate various phenomena in monolayer graphene such as Klein tunneling \cite{Katsnelson_2006, Cheianov_2006, Zhang_2008, Shytov_2008, Young_2009, Stander_2009}, electron collimation \cite{Campos2012,Rickhaus2013}, and the modulation of supercurrents \cite{Calado2015,Allen2017}. In addition, FP interference has been instrumental in characterizing layer-dependent electronic properties, revealing the transition between quantum and classical confinement in trilayer graphene~\cite{Campos2012}, anti-Klein tunneling in bilayer graphene \cite{Varlet2014, Du2018blg}, and topological channels\cite{Rickhaus2018} or electronic thickness\cite{Rickhaus2020, MrecaKolasiska2022} in twisted bilayer graphene. Moreover, FP resonances have proven to be a powerful probe for the rich physics of graphene-based superlattice cavities, unveiling phenomena such as moiré band reconstruction and anomalous cyclotron dynamics~\cite{Handschin2017, Kraft2020}. A more recent experimental effort has gone into building FP interferometers to probe anyonic statistics \cite{Halperin_2011, D_prez_2021, Werkmeister_2024}. Since FP interference is highly sensitive to cavity geometries and local electrostatics, understanding these effects is central to interpreting ballistic transport and designing interferometric architectures. In this work, we use FP interference as a non-invasive tool to probe the internal potential profile of a large-angle twisted bilayer graphene (tBLG) device, locating and sizing the unintentional doping that defines multiple electronic cavities. 

\begin{figure*}
    \includegraphics[width=0.8\textwidth]{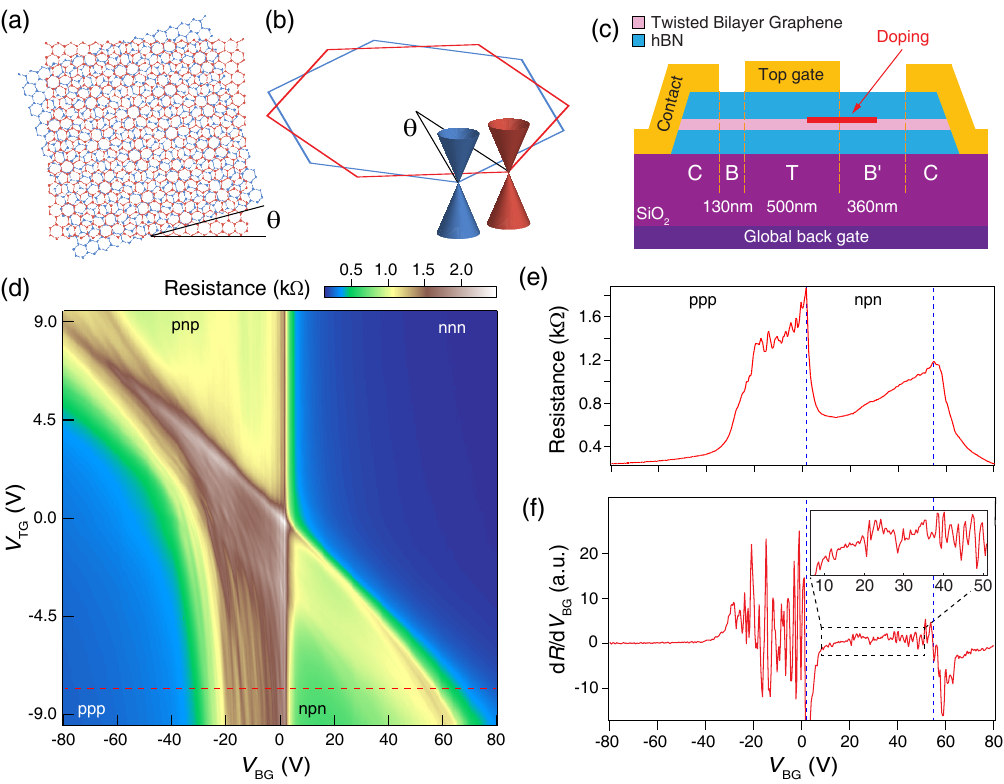}
  \caption{(a) Crystal structure of tBLG with a twist angle $\theta$. (b) Schematic low-energy band structure of the decoupled tBLG showing the separation of the Dirac cones by the same twist angle $\theta$ (c) Schematic diagram of our device. The global back gate and local top gate define the single gated regions, B and B$'$, and the dual gated region T. Regions underneath the contacts are labeled C. The thick red line indicates the location of the  unintentional local doping. (d) Resistance map as a function of $V_{\mathrm{BG}}$ and $V_{\mathrm{TG}}$ at $0$~T and $2.4$~K (e) Line cut of (d) along $V_{\mathrm{TG}}=-8V$ (f) Numerical derivative $dR/dV_{\mathrm{BG}}$ of (e); inset: zoom  highlighting faint oscillations observed in the npn region}
  \label{Fig. 1}
\end{figure*}

For large angle tBLG, the Dirac cones of the two layers are widely separated in $k$-space (Figure~\ref{Fig. 1}a-b), strongly suppressing inter-layer tunneling and yielding effectively decoupled layer-specific transport channels \cite{Luican_2011, Sanchez_Yamagishi_2012, Sanchez_Yamagishi_2016, Rickhaus2020, Kim_2022, MrecaKolasiska2022, Inbar_2023, Babich_2025}. This allows FP interferometers to form in parallel through each graphene layer within a single device despite a small interlayer distance. 

Our device is a large-angle tBLG encapsulated between hexagonal boron nitride with a global back gate and a local top gate defined lithographically, which divides the channel into single-gated regions B and B$'$ and a dual-gated region T between contact regions C (Figure~\ref{Fig. 1}c). The capacitive couplings of the back gate $C_{\mathrm{BG}}$ and the top gate $C_{\mathrm{TG}}$ are extracted from the Landau levels in the quantum Hall data (see Supplementary Material). The mobility reaches $\mu = 130,000$~$\mathrm{cm^{2}~V^{-1}~s^{-1}}$, corresponding to a mean free path of up to $\sim$$2$~$\mu$m which exceeds cavity lengths, putting transport in the ballistic regime. The total carrier densities in the single-gated (SG) and dual-gated (DG) regions are given by equations $n_{\mathrm{SG}} = (C_{\mathrm{BG}} V_{\mathrm{BG}})/e$ and $n_{\mathrm{DG}} = (C_{\mathrm{BG}} V_{\mathrm{BG}} + C_{\mathrm{TG}} V_{\mathrm{TG}})/e$, respectively, where $V_{\mathrm{BG}}$ ($V_{\mathrm{TG}}$) is the back (top) gate voltage and $e$ is the elementary charge. 

\section{Results and Discussion}

\subsection{Transport Measurements and Modeling}

A 2-probe resistance map at zero magnetic field, $B=0$~T, and low temperature, $T=2.4$~K, as a function of $V_\mathrm{BG}$ and $V_\mathrm{TG}$ is shown in Figure 1d, where primary vertical and diagonal ridges can be seen to divide the map into unipolar (ppp and nnn) and bipolar (pnp and npn) transport regimes.  These vertical and diagonal resistance peaks correspond to the charge neutrality points of the single-gated and the dual-gated regions, respectively. At high displacement fields, the diagonal Dirac peak splits into two distinct branches. This behavior is frequently observed in tBLG and is attributed to the vertical electric field lifting the layer degeneracy, thereby resolving the charge neutrality points of the individual layers\cite{Rickhaus2020, Slizovskiy_2021, MrecaKolasiska2022}.

In addition to the primary Dirac features, the measurement reveals complex fine structure in the resistance map. A weaker secondary vertical resistance peak is observed near $V_{\mathrm{BG}} \approx -20$~V, accompanied by another secondary diagonal peak shifted slightly below the main resistance ridge. Furthermore, in the region confined between the primary and secondary vertical peaks ($-20~\mathrm{V} < V_{\mathrm{BG}} < 0$~V) and at negative top-gate voltages ($V_{\mathrm{TG}} < 0$~V), we observe a series of pronounced vertical resistance oscillations.  

To further investigate these oscillations, Figure 1e shows the resistance profile along the cut indicated by the red dashed line ($V_{\mathrm{TG}} = -8$~V).  In the unipolar regime, the resistance oscillations are clearly visible on the shoulder of the main resistance peak ($-20~\mathrm{V} < V_{\mathrm{BG}} < 0$ V). To better resolve these features, we plot the numerical derivative $dR/dV_{\mathrm{BG}}$ in Figure 1f. This highlights the strong unipolar oscillations and also reveals  fine oscillations in the bipolar npn region, which is less apparent in the raw resistance data.

While resistance oscillations in the bipolar regime are characteristic of FP interference, their observation in the nominally unipolar regime is unexpected. This anomaly suggests that the ideal unipolar classification is disrupted by unintentional local doping. The resistance in this oscillatory region is unusually high, exceeding typical values for both unipolar and bipolar states. Such high resistance points to the formation of unplanned potential barriers. We attribute these features to localized, unintended doping which manifests as secondary Dirac peaks in Figure 1d. 

The secondary vertical peak depends exclusively on $V_{\mathrm{BG}}$, placing it in the single-gated region, whereas the secondary diagonal peak arises from doping within the dual-gated channel. From the voltage offsets of these secondary vertical and diagonal neutrality points, we extract carrier density shifts of $1.8\times10^{12}~\mathrm{cm^{-2}}$ and $ 1.6\times10^{12}~\mathrm{cm^{-2}}$, respectively. The comparable magnitudes of these shifts point to a common doping source spanning the boundary between the single-gated  and dual-gated regions. 

To determine the spatial extent of this doping, we compared transverse 2-probe measurements across multiple contact pairs along the device (see Supplementary Material). The consistent observation of similar secondary Dirac peaks and oscillation patterns indicates that the unintentional local doping possesses a highly correlated spatial distribution. Whether manifesting as a continuous feature along the top-gate edge or as localized islands at the intersections between the top gate and the graphene leads, its systematic nature strongly suggests a fabrication-related origin. Notably, this specific geometry explains how the high carrier mobility measured in our 4-probe configuration can coexist with such a substantial density step, because the longitudinal 4-probe current flows parallel to and avoids the barrier, whereas transverse 2-probe measurements directly capture it (see Supplementary Material for more details). By incorporating this localized doping profile into our device model (Figure~\ref{Fig. 1}c), we are able to qualitatively reproduce the oscillation modes observed in both the bipolar and unipolar regimes.

\begin{figure*}
  \centering
    \includegraphics[width=0.8\textwidth]{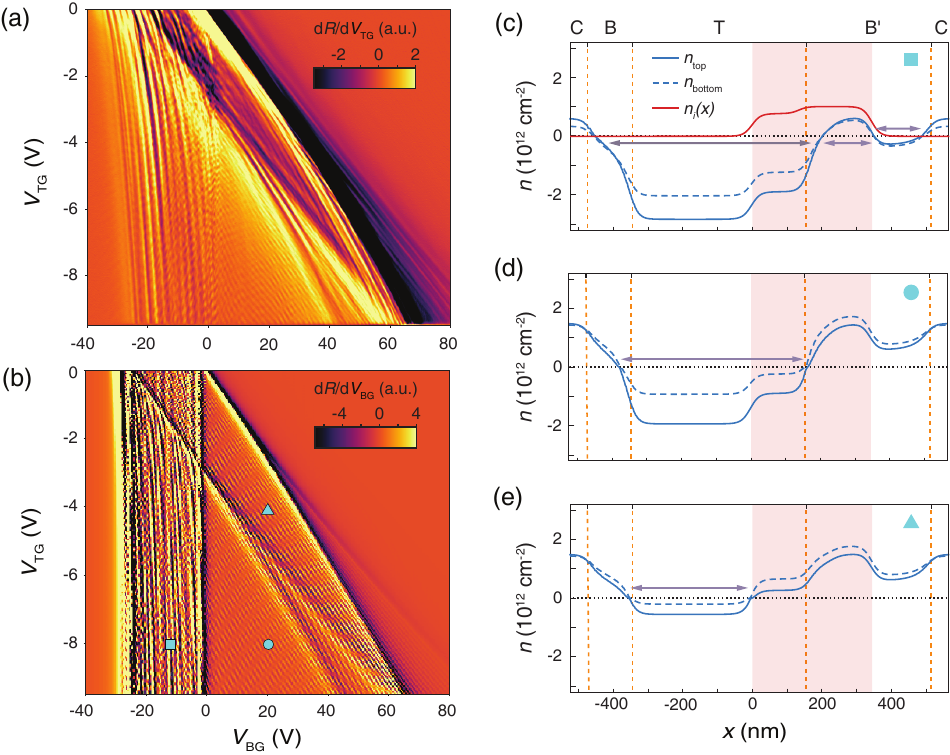}
  \caption{(a, b) Experimental map of the numerical derivative $dR/dV_{\mathrm{TG}}$ and simulated map of $dR/dV_{\mathrm{BG}}$ as a function of $V_{\mathrm{BG}}$ and $V_{\mathrm{TG}}$  at $B=0$ T and $T=2.4$ K. The simulation (b) reproduces the experimental patterns using the extracted local doping profile. Markers indicate gate configurations for panels (c-e).(c-e) Carrier density profiles corresponding to the square, circle, and triangle markers in (b). Solid and dashed blue lines denote top and bottom layer density profiles, respectively. The red line in (c) represents the unintentional local doping profile. Double-headed arrows mark the resonant cavities, highlighting (c) a series of cavities in the unipolar regime, (d) a long central cavity, and (e) a shortened cavity modified by the gate potential.}
  \label{Fig. 2}
\end{figure*}

To clearly resolve the fine structure of these oscillations, Figure 2a presents a high-resolution map of numerical derivative of the resistance, $dR/dV_{\mathrm{TG}}$, focusing on the ppp and npn regions from Figure 1d (see Supplementary Material for a full map). The splitting of both diagonal charge neutrality lines becomes resolvable at high displacement fields. In addition, the data reveal a striking interference pattern characterized by three distinct regimes: a series of sharp vertical fringes in the range $-20~\mathrm{V} < V_{\mathrm{BG}} < 0$~V, a complex set of diagonal and curved oscillations in the npn region, and another set of oscillations sandwiched between the primary and secondary diagonal charge neutrality lines.

We reproduce these transport signatures using quantum transport simulations based on the Landauer-Büttiker formalism within a scalable tight-binding approximation~\cite{Buettiker1986,Baranger1989,Datta1995,zwierzycki_calculating_2008, Liu2015, kolasinski_interference_2016, MrencaKolasinska2023, Chakraborti2024} (Figure 2b; see Supplementary Material). The potential profile is determined via a self-consistent quantum capacitance model, which treats the decoupled graphene layers as a parallel capacitor influenced by external gates~\cite{Liu2013, Rickhaus2020} and accounts for contact-induced n-doping~\cite{Wang2013,Du2018blg,Zhao2025-hz}. Crucially, by incorporating the local doping profile $n_i(x)$ derived from our electrostatic analysis (Figure 2c, red curve), the simulation achieves excellent agreement with the experimental data. It captures not only the quasi-periodic spacing and orientation of the vertical and diagonal fringes but also the relative amplitude contrast between modes, providing strong support for the additional cavity model. Notably, this extracted doping profile serves as a highly constrained model that consistently reproduces the complex transport features, rather than representing a strictly unique mathematical solution.

To determine the physical origin of the interference fringes, we analyze the carrier density profiles in three different gate configurations marked by the symbols in Figure 2b. Figure 2c corresponds to the nominally unipolar regime (square marker; $V_{\mathrm{BG}} = -10$ V, $V_{\mathrm{TG}} = -8$ V). Here, the doping $n_i(x)$ modifies the potential profile in the single-gated region B$'$. This results in the formation of three distinct FP cavities: a long  central p-type cavity ($L \approx 650$~nm) in the dual-gated region T, a short n-type cavity in region B$'$ and a short p-type cavity in region B$'$. The three cavities manifest as vertical oscillations originating from the short p and n-type cavities in region B$'$ and faint fine diagonal oscillations from the long central p-type cavity.  

By analyzing the observed fringe spacing, we extract the cavity dimensions using the condition for constructive interference, $k_F \cdot 2L = 2\pi N$, where $N$ is an integer. For the vertical fringes observed in the unipolar regime, this yields a length of $L \approx 150$~nm. The vertical orientation confirms that these modes are modulated solely by the back gate, placing them within the single-gated region B$'$. Since region B$'$ has an effective length of roughly 300~nm (Figure 2c), we conclude that it is partitioned into two adjacent segments—one n-type and one p-type—of comparable size ($L \approx 150$~nm). These series cavities generate the prominent vertical interference pattern that dominates the unipolar transport.

In the bipolar regime (Figure 2d, circle marker), the device forms a global npn junction. Despite the incorporation of the unintentional local doping, the effective cavity spans the full dual-gated region, generating the diagonal interference pattern typical of tBLG. Increasing the top-gate voltage (Figure 2e, triangle marker) drives the system across a secondary neutrality point, shortening the cavity and shifting the oscillation period. The difference in cavity length between these configurations corresponds to the doping segment directly under the top gate. We attribute the broader fringes observed in the triangle region to charge puddles forming at low carrier densities in the bottom layer~\cite{Xue2011, Decker2011}. The quantitative agreement between the measured and simulated oscillation periods across the majority of the gate voltage map demonstrates that transport spectroscopy can serve as a powerful, non-invasive probe for mapping the location and spatial extent of charge inhomogeneities in van der Waals heterostructures, even when they are buried within the device architecture.

Unlike conventional local probes that offer powerful real-space imaging but are physically restricted by encapsulating layers and metallic gates, this approach operates directly on fully completed devices. We achieve a spatial resolution on the tens-of-nanometers scale, a limit determined by intrinsic density fluctuations and by evaluating the sensitivity of our interference signatures to variations in the doping profile, including position, smoothness, and cavity length (see Supplementary Material). While this transport spectroscopy method generalizes to other 2D materials, the dual-layer nature of tBLG provides additional constraints for the analysis; the layers’ distinct capacitive couplings generate interference beating and Dirac peak splitting that provide rigorous constraints for the electrostatic model. Notably, this approach requires high-mobility devices with relatively coherent potential landscapes, as pervasive 2D charge puddle networks can dampen the necessary interference patterns.

\subsection{Magnetic Field Dependence}

\begin{figure*}
  \centering
        \includegraphics[width=0.9\textwidth]{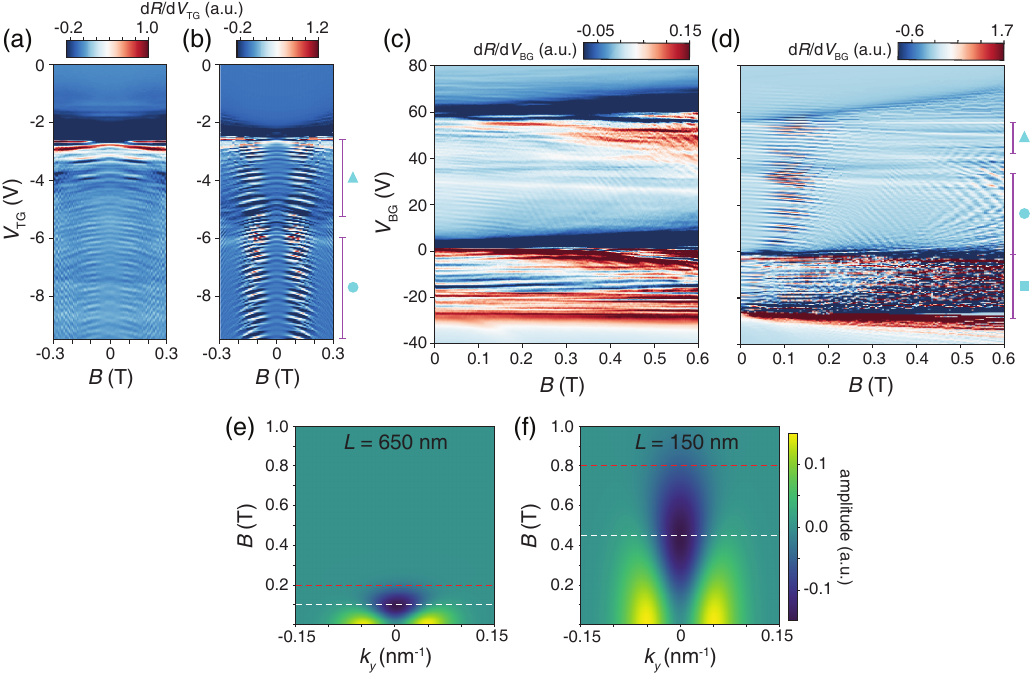}
  \caption{(a, b) Experimental and simulated maps of the numerical derivative $dR/dV_{\mathrm{TG}}$ versus $B$ in the bipolar regime ($V_{\mathrm{BG}} = 20$ V). The interference fringes disperse and vanish for $|B| > 0.2$ T due to the cyclotron cutoff.
(c, d) Experimental and simulated maps of $dR/dV_{\mathrm{BG}}$ versus $B$  at $V_{\mathrm{TG}} = -8$ V. The quasi-periodic fluctuations persist with weak dispersion to high fields, supporting confinement in a small, electronic cavity. (e, f) Calculated oscillation amplitude for (e) long ($L=650$~nm) and (f) short ($L=150$~nm) cavities. The model shows that the long-cavity mode is suppressed at low fields ($B \approx 0.2$~T), while the short-cavity mode survives well beyond the measurement range.}
  \label{Fig. 3}
\end{figure*}

To further verify the origin of the oscillations, we measured their magnetic field dependence. Figure 3a shows the evolution of the differential resistance $dR/dV_{\mathrm{TG}}$ with magnetic field $B$ at fixed $V_{\mathrm{BG}} = 20$~V (bipolar regime), compared with simulations in Figure 3b. Both datasets show a distinct bending of the interference fringes for $|B| < 0.2$~T. This dispersion reflects the Lorentz force curving electron trajectories , which alters the effective path length and interference phase within the cavity.

Figure 3c maps the magnetotransport at fixed $V_{\mathrm{TG}} = -8$~V, spanning both the  unipolar ($V_{\mathrm{BG}} < 0$~V) and bipolar ($V_{\mathrm{BG}} > 0$~V) regimes. The simulation (Figure 3d) captures the distinct behaviors in both regions using the single doping profile derived earlier. In the nominally unipolar range (square marker), the quasi-periodic fluctuations exhibit a subtle yet discernible magnetic dispersion,  indicating that they are coherent interference modes rather than noise. Simultaneously, the bipolar region displays the expected channel-wide resonances. This consistent agreement across the full voltage range provides strong evidence for the cavity model defined by the unintentional local doping.

The impact of cavity size on magnetotransport is clear when comparing the two regimes. In the long-cavity bipolar region (Figure 3a, $L \approx 650$~nm), the fringes disperse rapidly and vanish above $|B| \approx 0.2$~T. This cutoff occurs when the cyclotron radius $r_c = \hbar k_F / eB$ shrinks below the cavity length, bending electron trajectories enough to break the resonant condition. A sharp transition appears at $V_{\mathrm{TG}} \approx -5.5$~V. As the system crosses the secondary neutrality point and the effective cavity shortens (triangle marker, Figure 2e), the oscillations survive to higher fields. This robustness mirrors the unipolar regime (Figure 3c), where the  cavity is small ($L \approx 150$~nm). In these short-cavity limits, the cyclotron radius remains larger than the confinement length over a wider field range, preserving the ballistic paths necessary for interference.

To provide a quantitative understanding of these magnetic field scales, we calculate the theoretical oscillation amplitude as a function of  transverse momentum $k_y$ and magnetic field \cite{Young_2009} (see Supplementary Material for details). Figure 3e models the long cavity regime ($L = 650$~nm) corresponding to the bipolar junction. The model predicts a clear amplitude maximum at $\sim$$0.1$~T followed by suppression beyond $0.2$~T, both matching the experimental data. This peak arises from the competition between Klein tunneling and magnetic bending. At zero field, near-perfect transmission minimizes the reflection needed for strong FP interference. A weak magnetic field bends electron trajectories~\cite{Taychatanapat2013, Klanurak2024}, increasing reflection for off-normal modes and optimizing the interference visibility before the cyclotron cutoff eventually blocks transmission~\cite{Young_2009, Cheianov_2006, Shytov_2008}. Furthermore, a characteristic Klein tunneling $\pi$-phase shift is captured in both the experimental data and the electrostatic simulations. For example, in Figure 3a, this phase signature manifests at approximately 50 mT for the cavity mode corresponding to the circle region, and at roughly 70 mT for the triangle region (see Supplementary Material for details).

\begin{figure*}
  \centering
    \includegraphics[width=0.8\textwidth]{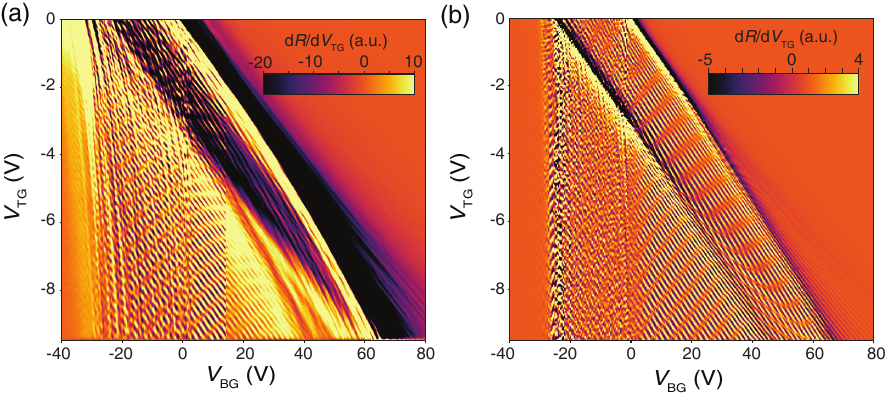}
  \caption{(a) Experimental map of the numerical derivative $dR/dV_{\mathrm{TG}}$ at $B = 0.1$ T. Diagonal oscillations from the long central cavity extend into the unipolar regime ($V_{\mathrm{BG}} < 0$ V), where they coexist with sharp vertical fringes. A switching event is observed at $V_{\mathrm{BG}} \approx 12$ V. (b) Simulated map of $dR/dV_{\mathrm{TG}}$ at $B = 0.1$ T. The model qualitatively reproduces the key signature of the three-cavity series: the superposition of dispersive diagonal modes and vertical fringes in the unipolar region.}
  \label{Fig. 4}
\end{figure*}

In contrast, the calculation for the short cavity ($L = 150$~nm) predicts that the amplitude peak is shifted to a much higher field of $0.45$~T, with the ballistic cutoff extending to $0.8$~T (Figure 3f). Although the predicted amplitude peak is not explicitly resolved in the experimental data (Figure 3c, square marker), likely due to the superposition of 3 different cavity modes in this regime, the simulation explains why the vertical fringes do not vanish. The theoretical cutoff of $0.8$~T lies beyond our measurement range ($B_{\mathrm{max}} = 0.6$~T), consistent with the experimental observation that the unipolar oscillations persist throughout the entire scanned magnetic field range.

To fully resolve the complex mode coexistence hypothesized in Figure 2c, we acquire high-resolution resistance maps at a fixed magnetic field of $B = 0.1$~T (Figure 4a, experiment; Figure 4b, simulation). We select this specific field magnitude because it corresponds to the maximum visibility of the FP oscillations in the long cavity in Figure 3a. This measurement reveals a feature obscured at zero field: diagonal oscillations extending deep into the nominally unipolar regime ($V_{\mathrm{BG}} < 0$~V). These diagonal fringes correspond to the long central cavity and are observed to coexist with the sharp vertical fringes. This superposition provides strong experiment evidence for the three-cavity configuration (Figure 2c), where the central p-type channel operates in series with the short n-type and p-type cavities in region B$'$. We note a device switching event at $V_{\mathrm{BG}} \approx 12$~V in the experimental data, likely caused by contact instability.

The simulation in Figure 4b does not fully capture the intricate complexity of the experimental interference pattern. We attribute this discrepancy to our simplified contact potential and the assumption  that the unintentional local doping is distributed equally between the two graphene layers (see Supplementary Material for details). Nevertheless, it successfully reproduces the coexistence of diagonal modes and vertical fringes in the unipolar region. This qualitative recovery of the long-cavity mode supports the view that the device operates as a series of cavities defined by the extracted local doping profile.

\section{Conclusion}

In summary, we utilize FP interference to probe the potential profile of dual-gated twisted bilayer graphene. By analyzing zero-field interference patterns, we identify and deduce the dimensions of a hidden series of unintentional doping-induced electronic cavities that partition the device. Magnetotransport measurements are consistent with this model, supporting the coherent nature of the modes and distinguishing localized resonances from the main channel signal. Our findings illustrate that complex oscillation patterns in nominally unipolar regimes can originate from unintentional local doping.

\begin{acknowledgments}
We thank Tan Jun You for assistance with device fabrication. Th.T. acknowledges support from the National Research Council of Thailand (NRCT) and Chulalongkorn University (Grant No. N42A650266); the NSRF via the Program Management Unit for Human Resources \& Institutional Development, Research and Innovation (Grant No. B39G680007); the Office of the Permanent Secretary of the Ministry of Higher Education, Science, Research and Innovation; and CU Power Grant, C2F. A.M.-K. acknowledges support from the National Science Centre, Poland (Project No. 2024/55/D/ST3/00538) and the Polish high-performance computing infrastructure PLGrid (HPC Center: ACK Cyfronet AGH) for providing computer facilities and support within computational Grant No. PLG/2025/018638. K.W. and Ta.T. acknowledge support from JSPS KAKENHI (Grant Nos. 21H05233 and 23H02052), the CREST (Grant No. JPMJCR24A5), JST and World Premier International Research Center Initiative (WPI), MEXT, Japan. M.-H.L. acknowledges financial support from the National Science and Technology Council (NSTC) of Taiwan (Grant Nos. 114-2112-M-006-029-MY3 and 112-2112-M-006-019-MY3).

Th.T. conceived and supervised the project. C.M. and Th.T. conducted the transport measurements. C.M., A.M.-K., and Th.T. performed the data analysis. A.M.-K. and M.-H.L. developed the theoretical model and performed the simulations. K.W. and Ta.T. provided the high-quality hBN crystals. C.M., A.M.-K., and Th.T. wrote the manuscript with input from M.-H.L. All authors reviewed the final results.

The simulation data underlying this study are openly available in the Rodbuk repository at https://doi.org/10.58032/AGH/6VHQL1
\end{acknowledgments}

%

\newpage
\onecolumngrid
\renewcommand{\theequation}{S\arabic{equation}}
\renewcommand{\thefigure}{S\arabic{figure}}
\renewcommand{\thetable}{S\arabic{table}}
\renewcommand{\theenumi}{S\arabic{enumi}}
\renewcommand{\thesection}{S\arabic{section}}
\setcounter{figure}{0} 
\setcounter{section}{0}

\begin{center}
	\LARGE{\bf Supplementary Material}
\end{center}

\section{Device Characteristics}

\subsection{Fabrication} 
\begin{figure}[tb]
    \includegraphics{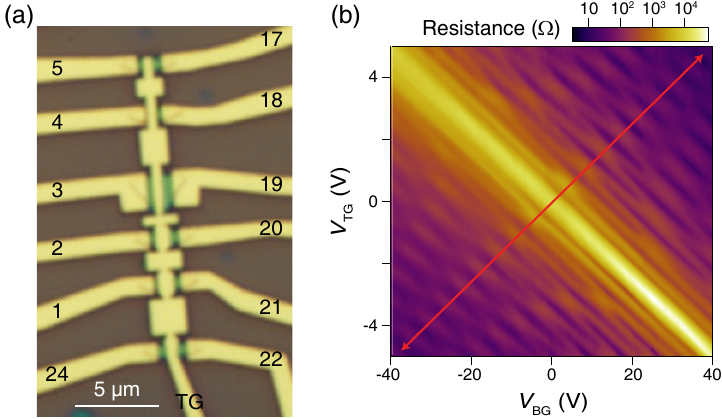}
    \caption{(a) Optical micrograph of the twisted bilayer graphene device. TG denotes the top gate, and gold contacts are numbered. Main text data were acquired between contacts 4 and 18. Scale bar: 5 $\mu$m. (b) Magnetotransport map of longitudinal resistance $R_{xx}$ at 3 T. The intersecting features emerge from top and bottom layer Landau levels tuned by the out-of-plane displacement field. The red arrow traces the $D=0$ axis, indicating equal charge filling across both layers.}
    \label{3T}
\end{figure}

Monolayer graphene and hexagonal boron nitride (hBN) flakes were exfoliated from bulk crystals using the scotch tape method \cite{shuang2015reliable} and deposited onto Si/SiO\textsubscript{2} substrates. Candidate flakes were identified by optical contrast \cite{swang2012thickness}, and from the measured top and back gate capacitances  ($C_\mathrm{TG}$ and $C_\mathrm{BG}$) we estimate the top and bottom hBN thicknesses to be $38$~nm and $32$~nm, respectively. To assemble the heterostructure, hBN was intentionally misaligned with graphene and used to tear a monolayer graphene which was then rotated by a finite twist angle and re-stacked to form a hBN/tBLG/hBN stack. The completed heterostructure was released onto a Si/SiO\textsubscript{2} chip at $180$~$^\circ$C to minimize trapped blisters. Electron beam lithography was subsequently employed to define the top gate and contact regions, followed by thermal evaporation of Cr/Au ($5~$nm/$100~$nm) to form electrical contacts and the top gate electrode (Figure~\ref{3T}a).

\subsection{Determining $C_\mathrm{TG}$ and $C_\mathrm{BG}$} 
\rev{At $B = 3$ T, we observe a clear diagonal crossing pattern of layer-polarized Landau levels (Figure S1b). This phenomenon is a well-known signature of large-angle tBLG that has been previously observed in similar devices\cite{sSanchez_Yamagishi_2012}. Because the two decoupled graphene layers act as independent channels, tuning the transverse displacement field induces an interlayer potential difference. This shifts their respective Dirac points in energy, driving the Landau levels to cross.}

The total carrier density corresponding to the Landau level of filling factor $\nu$ is given by $n_{tot} = \nu eB/h$. Along the $D=0$ line, the two graphene layers are equally filled, such that the total carrier density is related to the back gate voltage by $n_{tot} = 2 C_\mathrm{BG} V_\mathrm{BG} /e$. By assigning filling factors to multiple Landau level crossings and performing a linear fit of $n_{tot}$ versus $V_\mathrm{BG}$, we extract $C_\mathrm{BG} = 1.08\times 10^{-4}$ F/m$^2$. The slope of the $D=0$ line yields the capacitance ratio $C_\mathrm{BG}/C_\mathrm{TG}=0.128$, from which we obtain $C_\mathrm{TG} = 8.87\times 10^{-4}$ F/m$^2$.

\section{Transport Data}
  
\subsection{Local transport measurements} \rev{Measurements were performed using standard low-frequency lock-in techniques operated at a reference frequency of 17 Hz. Excitation currents ranging from 100 nA to 1 $\mu$A were used, and all measurements were carried out at 2.4 K. Specifically, the two-probe data presented in Figure 1 of the main text were measured between contacts 4 and 18. In contrast, the longitudinal resistance ($R_{xx}$) data in Figure S1b were obtained using a four-probe configuration located exclusively on one side of the sample, with current injected between contacts 22 and 17 and voltage probed across contacts 21 and 20 (see Figure S1a for contact numbering).}

\rev{The high mobility and pristine Shubnikov–de Haas (SdH) oscillations observed in our standard 4-probe measurements result directly from the measurement geometry.  Because the longitudinal contacts are aligned parallel to the unintentional local doping, the $\Delta n \sim 10^{12}~\mathrm{cm}^{-2}$ density barrier contributes no series resistance, allowing the extracted zero-field mobility to reflect the pristine hBN-encapsulated bulk. Furthermore, at high magnetic fields, the clean SdH oscillations stem from this geometry and quantum transport dynamics. Specifically, when the Fermi level lies inside a Landau level, the longitudinal voltage is governed by the bulk transport path. Conversely, when it resides in the gap, transport is mediated by edge states where backscattering is suppressed, as the edge disorder does not span the bulk to couple counter-propagating states at opposite edges. As a result, our longitudinal 4-probe configuration yields pristine SdH oscillations. In contrast, 2-probe measurements taken transversely across the sample force electrons to traverse the macroscopic potential barrier, which manifests directly as the secondary Dirac peaks in Figures 1d and 3c ($V_{\mathrm{bg}} \approx 30$~V).}

\subsection{Derivative of zero field data} 
\begin{figure}[tb]
    \includegraphics{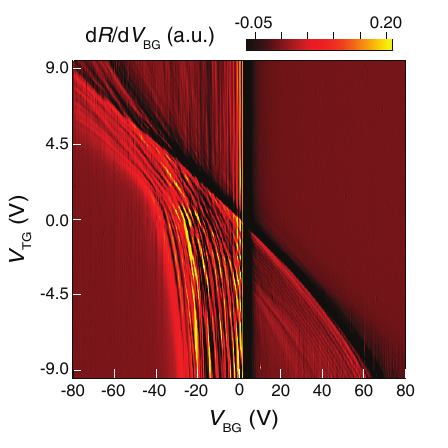}
    \caption{Resistance differentiated with respect to $V_\mathrm{BG}$ of Figure 1d in the main text.}
    \label{diff0T}
\end{figure}

Figure~\ref{diff0T} shows the derivative with respect to $V_\mathrm{BG}$ for the experimental data from Figure 1d in the main text. The vertical oscillations are clearly resolved and are attributed to unintentional doping in the single gated region, which modulates the carrier density outside the top gate region. We further observe an asymmetry between the two bipolar regimes, npn and pnp. While FP interference patterns can be resolved in the npn region with higher-resolution maps (Figure 2 in the main text), no comparably clear FP oscillations are observed in the pnp regime. This asymmetry can be understood from the presence of intrinsically n-doped contacts which effectively transform the pnp configuration into an n-pnp-n configuration. The resulting additional p-n interfaces lead to multiple partially reflecting boundaries, increasing mode mixing, and reducing the phase coherence of charge carriers within the main cavity.

\begin{figure}[tb]
    \includegraphics{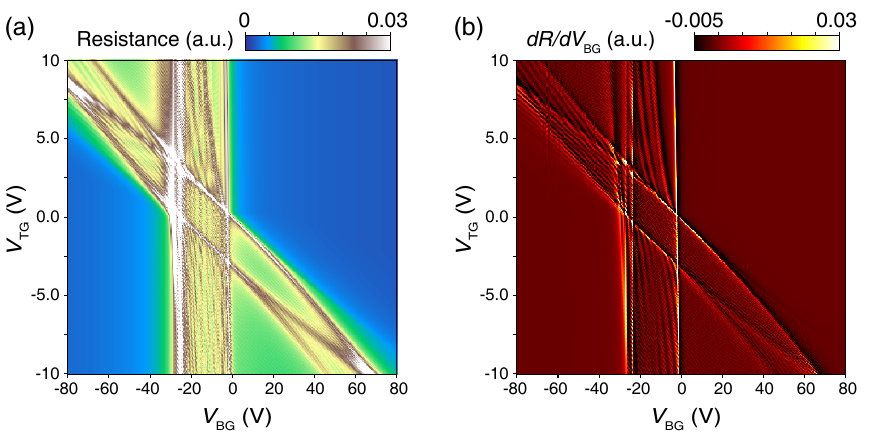}
    \caption{(a) Simulation of  resistance and (b) derivative of resistance at zero magnetic field for full top gate and back gate range, showing oscillations matching that of Fig. 1d in the main text.}
    \label{fullsim}
\end{figure}

\section{Spatial Distribution and Possible Origins of the Unintentional Local Doping} 

\begin{figure}[tb]
    \includegraphics{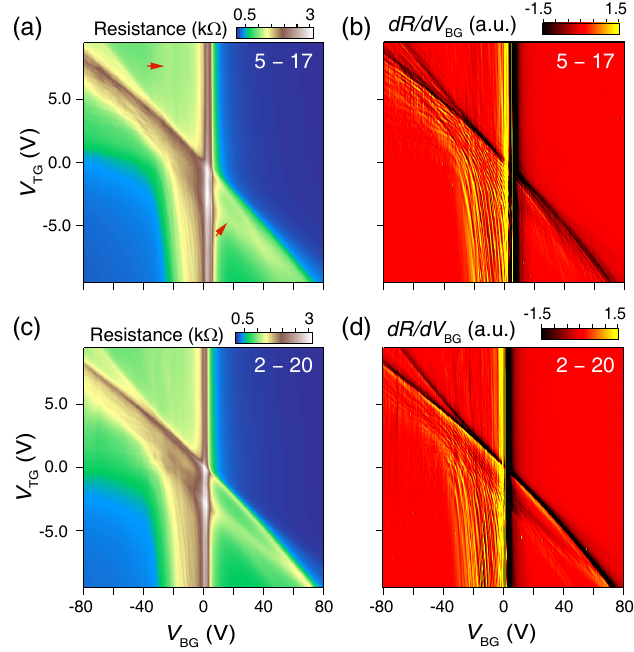}
    \caption{ (a–d) Contact pair consistency. The 2-probe resistance maps and corresponding $dR/dV_{\text{BG}}$ derivatives for (a, b) contact pair 5–17 and (c, d) contact pair 2–20. Both contact configurations display the same secondary Dirac peaks and Fabry-Pérot oscillations shown in the main text, \revv{suggesting} that the \rev{unintentional local doping} extends macroscopically along the edge of the top gate.}
    \label{twoprobe}
\end{figure}

The 2-probe measurements acquired across adjacent contact pairs 5–17 (Figure~\ref{twoprobe}a, b) and 2–20 (Figure~\ref{twoprobe}c, d) are compared with the main-text data taken between contacts 4 and 18. All three contact pairs consistently exhibit similar secondary Dirac peaks (red arrows in Figure~\ref{twoprobe}a) and Fabry-Pérot interference patterns. The repeated observation of these features indicates that the unintentional local doping is not a random, isolated point defect, but rather possesses a highly correlated spatial distribution.

\rev{Whether this unintentional local doping manifests as a continuous feature along the top-gate edge or as multiple localized islands at the intersections between the top gate and the graphene leads, its systematic nature strongly suggests a fabrication-related origin. Plausible causes include shadowing effects during metal evaporation, where fabrication residues or trapped charges accumulate near the resist boundary, localized strain relaxation during thermal cycling, or contact-induced doping and edge charge accumulation resulting from device etching. Because a random 2D network of charge puddles would destroy phase coherence, these mechanisms must manifest as a highly correlated 1D step to preserve the interference.}

\rev{Furthermore, mechanisms such as gate-fringe fields and layer-asymmetric electrostatics inherently exist in dual-gated devices and were explicitly included in our electrostatic simulations. However, the simulations reveal that these fringe fields cannot independently reproduce the vertical interference pattern observed in the nominal unipolar regime. Similarly, while out-of-plane layer asymmetry accounts for the observed beating patterns and Dirac peak splitting, it lacks the in-plane spatial confinement necessary to form a cavity.}

Regardless of the exact microscopic mechanism, the robust spatial correlation across multiple contacts and the quantitative agreement with our interference model \revv{strongly} point to an extended hidden cavity. This well-defined spatial correlation clearly distinguishes the feature from random bulk substrate defects, \revv{supporting its identification} as a systematic, albeit unintentional, structural partition of the device.

\section{Computational Details}

\subsection{Carrier densities in individual gated monolayer graphene}
Here we describe the quantum capacitance model in graphene in the presence of $N$ electrostatic gates and non-zero intrinsic doping $n_i$.
The carrier density in the zero temperature limit and zero $B$ is given by 
\begin{equation}
n(E) = \sgn(E)\frac{1}{\pi}\left(\frac{E}{\hbar v_F}\right)^2, 
\label{eq:density_B0}
\end{equation} 
If there is no external gating, but an intrinsic doping $n_i$ is present, the
quasi-Fermi level $E_i$ is described by 
 \begin{eqnarray}
n_i = \sgn(E_i)\frac{1}{\pi}\left(\frac{E_i}{\hbar v_F}\right)^2
\end{eqnarray} 
from which we can get $E_i$.
 
For obtaining the carrier density $n_G$ and electrostatic potential $V_G$ in the presence of external gates, we describe the carrier density in terms of the self-partial capacitances \cite{sLiu2013}. 
The net electron number density induced by gates on graphene is 
\begin{equation}
\label{eq:VG_general}
n_G = \frac{\rho_G}{-e} = \sum\limits_{j=1}^N  \frac{C_{jG}}{e} (V_j - V_G) ,
\end{equation} 
 where $C_{jG}$ is the self-partial capacitance of the $j$th gate. 
 If there is a non-zero intrinsic charge density $n_i$ on graphene, the
net carrier density of graphene is given by $n=n_i + n_G$, which should equal the density given by Eq.~(\ref{eq:density_B0}), i.e. $n_i + n_G = n(E_i + eV_G)$.
Thus we obtain 
\begin{equation}
\label{eq:VG_B0}
 \sum\limits_{j=1}^N \frac{C_{jG}}{e} (V_j - V_G) + n_i = \sgn(E_i + eV_G)\frac{1}{\pi}\left(\frac{E_i + eV_G}{\hbar v_F}\right)^2.
\end{equation} 
Here the $V_G$ value can be found numerically. 
The capacitances $C_{jG}$ are obtained from finite element electrostatic simulation, following Ref.\ \citenum{sLiu2013}. In general, they can be position-dependent, $C_{jG}=C_{jG}(x)$ and we can apply the same methodology for calculation of $V_G(x)$ at each $x$ point.
The model with linear dispersion relation described here can also be used at weak finite magnetic fields.

\subsection{Self-consistent method for capacitively coupled graphene layers}
When considering two large-angle decoupled graphene monolayers, we can treat them as an infinite parallel capacitor. Then, their electrostatic coupling is described in terms of the capacitance $C_G = \epsilon_0 \epsilon_G/d_G$, where $\epsilon_0$ is the permittivity of free space and we assume $\epsilon_G=1$ and  $d_G=0.12$ nm for interlayer distance\cite{sRickhaus2020electronicthickness}, resulting in $C_G/e = 460.53 \times 10^{15}$ V$^{-1}$m$^{-2}$. 
We refer to the electrostatic potential on the top (bottom) graphene layer as $V_{Gt}$ ($V_{Gb}$). These electric potentials can be treated as effective gate voltages with which the graphene monolayers influence each other. For example, the top graphene layer effectively feels a global back gate at a voltage $V_{Gb}$. Similarly, the bottom graphene layer effective feels a global top gate at a voltage $V_{Gt}$. 
Then, Eq.\ (\ref{eq:VG_B0}) for each graphene monolayer takes the form (assuming equal intrinsic doping $n_i$ on both layers)
\begin{align}
\label{eq:VGt_B0}
 \sum\limits_{j=1}^{N-1} \frac{C_{j,Gt}}{e} (V_j - V_{Gt}) +  \frac{C_{G}}{e} (V_{Gb} - V_{Gt}) + n_i &= \sgn(E_i + eV_{Gt})\frac{1}{\pi}\left(\frac{E_i + eV_{Gt}}{\hbar v_F}\right)^2, \\
\label{eq:VGb_B0}  \sum\limits_{j=1}^{N-1} \frac{C_{j,Gb}}{e} (V_j - V_{Gb}) +  \frac{C_{G}}{e} (V_{Gt} - V_{Gb}) + n_i &= \sgn(E_i + eV_{Gb})\frac{1}{\pi}\left(\frac{E_i + eV_{Gb}}{\hbar v_F}\right)^2.
\end{align} 
The gate voltages $V_j$ are input parameters; however, $V_{Gt}$ and $V_{Gb}$ are unknown, and they depend on each other. Therefore they are calculated self-consistently by iteratively solving equations (\ref{eq:VGt_B0}) and (\ref{eq:VGb_B0}) until convergence of $V_{Gt}$ and $V_{Gb}$ is achieved.

\subsection{Details of the electrostatic model}
\begin{figure}[tb]
\includegraphics[width=0.75\textwidth]{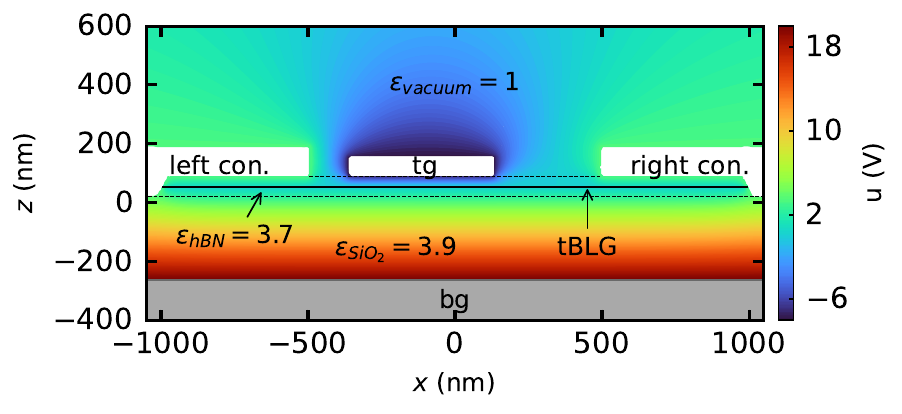}
	\caption{Example electrostatic potential calculated at $V_{C}=3.35$ V, $V_\mathrm{TG}=-8$ V, and $V_\mathrm{BG}=20$ V. Black solid line indicates the tBLG, and the dashed lines the hBN embedding it. }
	\label{fig:FigS1}
\end{figure}

Figure~\ref{fig:FigS1} shows the side view of the considered system, where the back gate at voltage $V_\mathrm{BG}$ is colored in gray and the top gate at voltage $V_\mathrm{TG}$, and contacts at  voltage $V_{C}$ are in white. The back gate is separated from the tBLG by a 285 nm thick SiO$_2$ with $\varepsilon_{SiO_2}=3.9$ and $d_{hBN,b} =32$ nm of hBN with $\varepsilon_{hBN}=3.7$, and the tBLG is covered by $d_{hBN,t} =38$ nm layer of hBN. The dimensions of the device follow Fig.\ 1(c) in the main text. The colormap shows an exemplary electrostatic potential map calculated at $V_{C}=3.35$ V, $V_\mathrm{TG}=-8$ V, and $V_\mathrm{BG}=20$ V. For this specific system, the top graphene monolayer is gated by the bottom layer, the local top gate (with the capacitive coupling described by the capacitance $C_{TG,Gt}$) and the contacts (described by $C_{C,Gt}$). 
For the bottom monolayer, we include the capacitive coupling to the other layer as well as the back gate  and contacts (described by $C_{BG,Gb}$ and $C_{C,Gb}$, respectively). 
Equations (\ref{eq:VGt_B0})--(\ref{eq:VGb_B0}) then become
\begin{align}
\frac{C_{TG,Gt}}{e} (V_\mathrm{TG} - V_{Gt}) + \frac{C_{C,Gt}}{e} (V_{C} - V_{Gt}) +  \frac{C_{G}}{e} (V_{Gb} - V_{Gt}) + n_i &= \sgn(E_i + eV_{Gt})\frac{1}{\pi}\left(\frac{E_i + eV_{Gt}}{\hbar v_F}\right)^2, \\
\frac{C_{BG,Gb}}{e} (V_\mathrm{BG} - V_{Gb}) + \frac{C_{C,Gb}}{e} (V_{C} - V_{Gb}) +  \frac{C_{G}}{e} (V_{Gt} - V_{Gb}) + n_i &= \sgn(E_i + eV_{Gb})\frac{1}{\pi}\left(\frac{E_i + eV_{Gb}}{\hbar v_F}\right)^2.
\end{align} 
For the calculation of capacitances $C_{BG,Gb}(x)$, $C_{C,Gb}(x)$, $C_{TG,Gt}(x)$, and $C_{C,Gt}(x)$ we use the approach in Ref.\ \citenum{sLiu2013}. 

The intrinsic doping $n_i$ is chosen to be equal on both graphene layers, and is described by a simple non-uniform density profile localized at the red region in Fig.\ 2(c) in the main text:
\begin{equation}
n_i(x) = \frac{n_L}{2}\left(\tanh\frac{x-x_1}{d_{s}} + \tanh\frac{-x+x_2}{d_s}\right)  + \frac{n_R}{2}\left(\tanh\frac{x-x_2}{d_s} + \tanh\frac{-x+x_3}{d_s}\right).
\end{equation} 
Here $d_s=100$ nm is the smoothness parameter, and the positions parameters are set to be $x_1=0$, $x_2=155$ nm and $x_3=345$ nm, leading to a good match between the simulated and the experimentally observed Fabry-P\'erot fringes (see Fig.\ 2(c) in the main text). The density in the dual-gated region $n_L$ and the back-gated region $n_R$ were optimized so that the charge neutrality lines in the simulation are as close as possible to our experimentally observed counterparts. This results in $n_L=0.8\times 10^{12}$ cm$^{-2}$ and $n_R=1\times 10^{12}$ cm$^{-2}$ on each layer, hence the total doping magnitude summed over the two layers is $2n_L=1.6\times 10^{12}$ cm$^{-2}$ and $2n_R=2\times 10^{12}$ cm$^{-2}$. This is close to the experimental estimation of $1.6\times 10^{12}$ cm$^{-2}$ in the dual-gated region and $1.8\times 10^{12}$ cm$^{-2}$ in the single-gated region; in the latter the small discrepancy may be because in the single-gated region the density is effectively lower due to the smoothness of the doping profile. 

Additionally, the contacts induce n-type doping in graphene; in order to include this in the electrostatic model we assume an effective contact voltage $V_C$ \cite{sDu2018blg}, which induces a density offset in the regions C shown in Fig.\ 2(c)--2(e) in the main text. Note that this doping, together with $n_i$, effectively extends the cavity underneath the top gate (see Fig.\ 2(c) in the main text), causing the Fabry-P\'erot oscillations to overlap with the nearly vertical fringes, especially seen in Fig.\ 4(a) of the main text. Therefore, for the estimation of contact parameter $V_C$, we chose the densities in the regions C to be nearly the density in B$^\prime$, so that $1.8\times 10^{12}$ cm$^{-2} = V_C C_C/e$ with $C_C/e = \epsilon_0\epsilon_{hBN} / ed_{hBN,b} = 0.531\times 10^{12}$ cm$^{-2}$, resulting in $V_C=3.35$ V. 

\subsection{Quantum transport}
The Hamiltonian of the system described in the tight-binding approximation is
\begin{equation}
\label{eq:Htb}
H = -\sum\limits_{\langle i, j\rangle} t_{ij} c_i^\dagger c_j + \sum\limits_i V(x_i, y_i) c_i^\dagger c_i,
\end{equation}
where in the first sum $\langle i, j\rangle$ runs over every nearest neighbor pairs, $t_{ij}$ is the hopping energy between the sites $i$ and $j$, $c_i^\dagger\ (c_i)$ is the creation (annihilation) operator of an electron at site $i$ located at a position $\mathbf{r}_i=(x_i, y_i)$, $V(x_i, y_i)$ is the onsite energy of the electron at site $i$. 
For realistic samples, to reduce the computational load we use the scalable tight-binding model for monolayer graphene \cite{sLiu2015}. In this model, the lattice spacing is $a=a_0 s_f$, where $a_0=0.142$ nm is the lattice spacing of the unscaled graphene, and $s_f$ is the scaling factor. At the same time, the nearest neighbor hopping parameter $t = t_0/s_f$, with $t_0=3$ eV, such that the low-energy band structure is kept invariant. In the presence of transverse magnetic field $\mathbf{B} = (0,0,B)$ the hopping integral becomes $t_{ij} = t \exp(i\phi)$, where we include the Peierls phase $\phi=(-e/\hbar) \int_{\textbf{r}_i}^{\textbf{r}_j} \textbf{A}\cdot d\textbf{r}$, and $\textbf{A}$ is the vector potential that satisfies $\nabla\times \mathbf{A} = \mathbf{B}$. 
In a multi-terminal system, we solve the scattering problem between all pairs of leads using the wavefunction matching method \cite{szwierzycki_calculating_2008, skolasinski_interference_2016}. Then the conductance is obtained from the Landauer-B\"uttiker formula $G_{kl}= 2(e^2/h) T_{lk}$, where $T_{lk}$ is the transmission probability from lead $k$ to lead $l$. Further, we construct the $N\times N$ conductance matrix $\mathbfcal{G}$, with elements given by \cite{sBuettiker1986, sDatta1995}
\begin{align}
\mathcal{G}_{kl} &= -G_{kl},\ \quad k\ne l, \\
\mathcal{G}_{kk} &= \sum\limits_{l=1,l\ne i}^N G_{kl},
\end{align}
where $N$ is the number of leads. 
The current $I_i$ in the $i$th lead is related to the voltage $V_j$ in the $j$th lead by $I_i = \sum_{j} \mathcal{G}_{ij} V_j$. The currents are related by the Kirchhoff's law, so we can set one of the voltages, say, in the $l$th lead to zero, and reduce the size of $\mathbfcal{G}$ to $(N-1)\times (N-1)$. Finally, it is inverted to obtain $\mathbfcal{R} = \mathbfcal{G}^{-1}$. Then, the 4-point resistance is
\begin{equation}
\label{eq:Rklmn} R_{kl,mn} = \frac{V_{m} - V_{n}}{I_k}, 
\end{equation}
where current flows from lead $k$ to $l$, and the voltage drop is taken between lead $m$ and $n$.

\subsection{The model system for the transport calculations}

\begin{figure}[tb]
\includegraphics[width=0.6\textwidth]{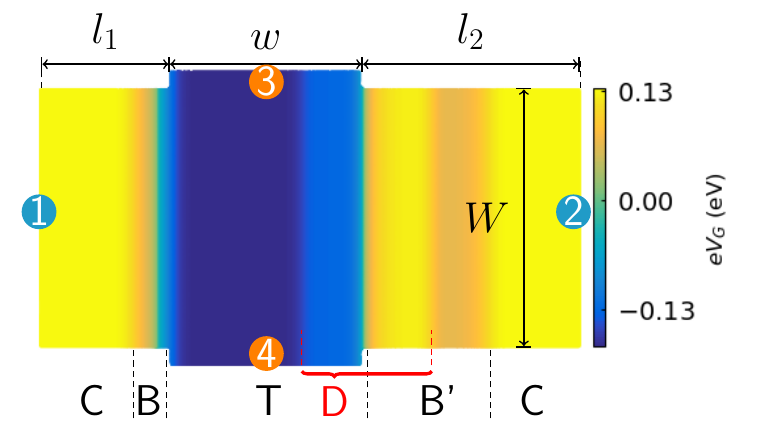}
	\caption{System geometry used in the transport calculation with onsite energy obtained at $V_\mathrm{TG}=-8$ V, and $V_\mathrm{BG}=10$ V. }
	\label{fig:FigS2}
\end{figure}
The system geometry is shown in Fig.~\ref{fig:FigS2} with an exemplary onsite energy map calculated at $V_\mathrm{TG}=-8$ V, and $V_\mathrm{BG}=10$ V, and the leads labeled by numbers 1--4. Here, leads labeled by 1 and 2 are voltage probes, and leads 3 and 4 are current probes. Resistance is calculated as
\begin{equation}
R_{34,12} = \frac{V_{2} - V_{1}}{I_3} = \mathcal{R}_{23} - \mathcal{R}_{13}. 
\end{equation}
The dimensions are chosen as $w=514$ nm, $W=697$ nm, $l_1=353$ nm, and $l_2=595$ nm. Within the scalable tight binding model we used the scaling factor $s_f=8$. To maintain the translational invariance in all the leads, we used the gauge transformation for $\mathbf{A}$ following Ref.\ \citenum{sBaranger1989, sMrencaKolasinska2023}. In the transport modeling we account for the contacts just by including the flat n-doped area (in the C regions) to simplify the geometry and reduce the computational cost. The exact shape of the edge contacts does not influence the Fabry-P\'erot oscillations in the lower quadrants of the $R(V_\mathrm{BG}, V_\mathrm{TG})$ relation. 

For the zero magnetic field resistance scan, we use the method of periodic boundary hopping \cite{sChakraborti2024}, where instead of a 4-terminal system, we assume a 2-terminal system which is translationally invariant along the lateral direction. Then, conductance is calculated as $G = (W/3\pi s_f a_{CC} )(g_b + g_t)$, where $g_j = (e^2/h)\int_{-k_F}^{k_F} T(k_y)dk_y$, $j=t,b$ labels the top (bottom) monolayer, $W=700$ nm, and $k_F$ is the Fermi momentum. Finally, resistance is calculated as inverse conductance $R=G^{-1}$.



\section{Sensitivity Analysis and Spatial Resolution Limits}

\begin{figure}
\includegraphics[width=0.85\textwidth]{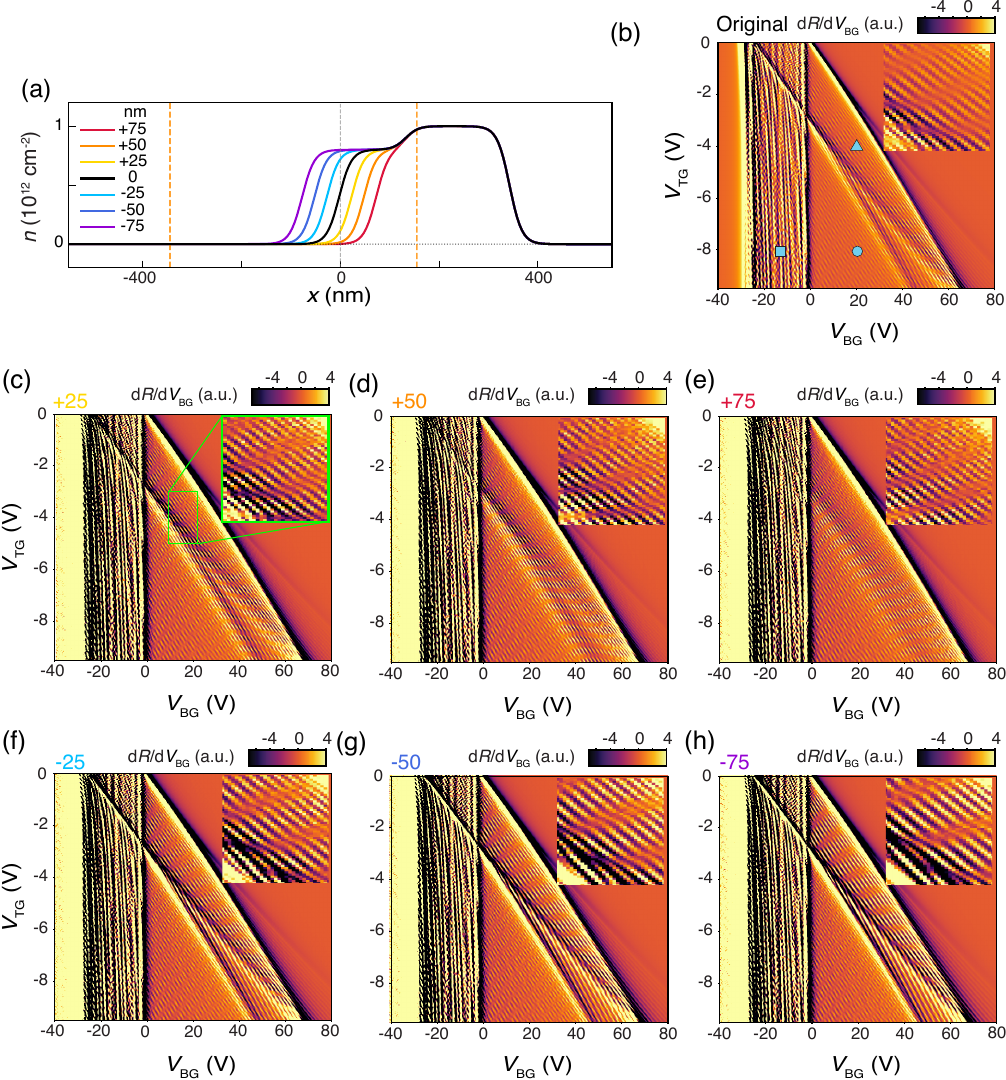}
	\caption{Effect of shifting the left boundary of the unintentional local doping profile. (a) Unintentional local doping profiles illustrating left-boundary shifts from -75 nm to +75 nm. (b) Original simulated interference pattern. (c-h) Derivative resistance maps for the shifted profiles, with insets magnifying the triangle region. Leftward shifts (negative values) strengthen the diagonal secondary Dirac peak and increase the interference period in the triangle region.}
	\label{fig:left_shift}
\end{figure}
\rev{To systematically evaluate how the specific geometry of the unintentional local doping governs the transport features, we simulate the effect of shifting its left spatial boundary while keeping the right boundary fixed (Figure~\ref{fig:left_shift}). Shifting this left border yields several distinct, localized effects on the interference patterns. First, as the border shifts further to the left (Figure~\ref{fig:left_shift}c–h), the total area encompassing the local doping increases, which directly enhances the intensity of the diagonal secondary Dirac peak. Second, the spatial shift heavily modulates the interference periods in the region denoted by the triangle marker. Because this specific Fabry-Pérot cavity is physically bounded by the left edge of the top gate and the left border of the local doping profile, shifting the doping boundary to the left reduces the effective cavity length. This reduction in cavity length consequently leads to a distinct increase in the oscillation periods. In contrast, the oscillation periods in the circle region remain completely unaffected; these fringes originate from the n-p-n cavity defined strictly by the physical width of the top gate, which remains constant. Similarly, the periods of the vertical fringes are strictly invariant, as the length of that cavity is dictated by the distance between the right edge of the top gate and the right border of the local doping profile, neither of which are altered in this simulation.}
\begin{figure}
\includegraphics[width=0.75\textwidth]{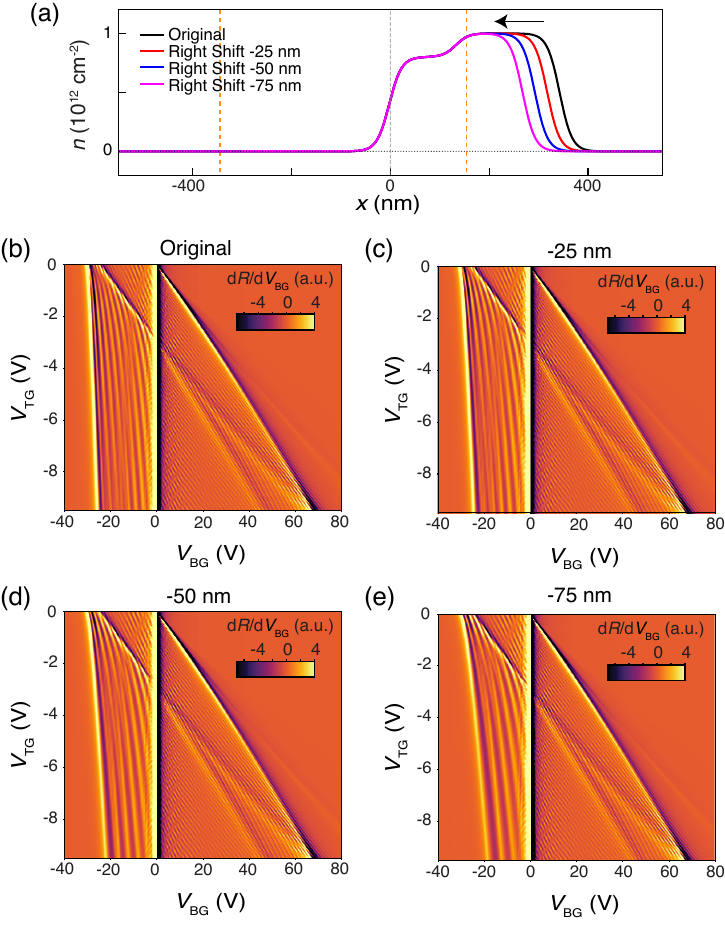}
	\caption{Effect of shifting the right boundary of the unintentional local doping profile. (a) Simulated unintentional local doping profiles without contact-induced n-doping, illustrating leftward shifts of the right doping boundary from -25 nm to -75 nm. (b-e) Corresponding simulated derivative resistance maps. Shifting the right boundary closer to the top gate visibly increases the oscillation period of the vertical fringes due to the reduced cavity length.}
	\label{fig:right_shift}
\end{figure}

\rev{In a complementary analysis, we investigate the right boundary of the unintentional local doping profile by simulating the interference patterns as this border is shifted leftward, closer to the top gate (Figure~\ref{fig:right_shift}). Because this shift does not alter the physical dimensions of the cavities defining the circle and triangle regions, their oscillation periods remain unaffected. Instead, the spatial shift exclusively modulates the vertical interference fringes located in the square region. In the full device model, the interference in this regime is highly convoluted due to the superposition of three distinct cavity modes (Figure 2c). To observe the boundary shift effect more clearly, we remove the contact-induced n-doping from this specific simulation. This simplifies the system to a single cavity mode, bounded by the right edge of the top gate and the right border of the local doping profile. As shown in the derivative resistance maps, moving the right boundary closer to the top gate decreases the effective cavity length, which directly manifests as a systematic increase in the oscillation period of the vertical fringes.}

\begin{figure}[tb]
\includegraphics[width=\textwidth]{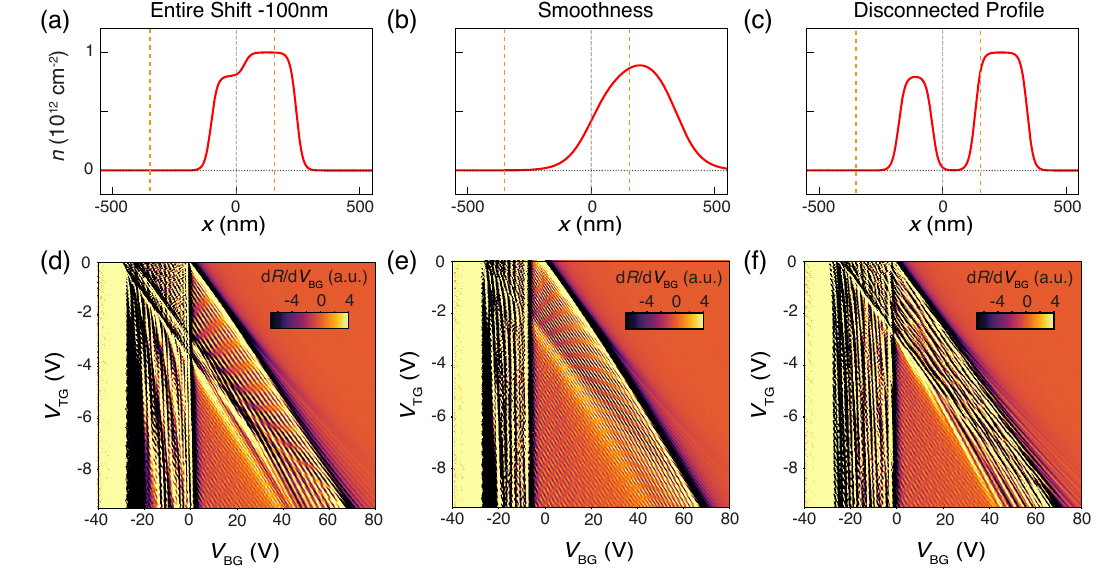}
	\caption{(a) Unintentional local doping profile and (d) corresponding derivative resistance map for a rigid 100 nm leftward shift of the entire doped region. (b, e) Simulation illustrating the effect of a heavily smoothed doping profile. (c, f) Simulation of a disconnected doping profile consisting of two separate spatial sites.}
	\label{fig:robus}
\end{figure}
\rev{Finally, we examine three further geometric modifications to the unintentional local doping profile to complete our systematic evaluation. First, we apply a rigid leftward shift of 100 nm to the entire doping region (Figure S9a, d). As expected, this global shift combines the effects of the individual boundary adjustments discussed previously. Because the left and right borders of the doping profile move closer to the left and right edges of the top gate, respectively, the corresponding cavity lengths decrease. This reduction systematically increases the oscillation periods in both the triangle and square regions. Next, we evaluate the effect of modifying the profile's edge smoothness (Figure S9b, e). When the doping profile is highly smooth, the simulated interference fringes connect continuously across the boundary between the circle and triangle regions. This directly contrasts with the original profile, where the sharper potential step creates abrupt discontinuities in the interference patterns as the system transitions between these regions. Lastly, we model a disconnected doping profile comprising two distinct spatial sites (Figure S9c, f). Although this split configuration still accurately reproduces the macroscopic positions of the secondary Dirac peaks, the resulting interference pattern, particularly within the triangle region, differs drastically from our experimental observations, providing an additional constraint on the physical geometry of the doping in the actual device.}

\rev{Beyond the geometric sensitivity of the interference patterns evaluated above, the spatial resolution ($\Delta L$) of this method is constrained by how accurately we can measure the interference period, which is inherently limited by intrinsic density fluctuations ($\delta n$). The Fabry-Pérot resonance condition dictates that the round-trip phase accumulation is $2k_FL = 2\pi j$ (where $j$ is an integer). Substituting the graphene Fermi wavevector $k_F = \sqrt{\pi n}$, we find that the extracted cavity length is $L = j\sqrt{\pi/n}$. Because the length scales with density as $L \propto n^{-1/2}$, the spatial resolution limit is roughly given by the derivative relation $\Delta L/L \sim \delta n/2n$. Given the intrinsic density fluctuations ($\delta n \sim 10^{11}\text{ cm}^{-2}$) at a typical operating density of $n \sim 10^{12}\text{ cm}^{-2}$, the relative uncertainty in the extracted cavity length is approximately 5\%. For the cavities in our devices, this translates to an uncertainty of $\Delta L \approx 10$--$30$~nm. Therefore, both our geometric simulations and theoretical error boundaries \revv{support the conclusion} that this methodology provides a highly robust framework for resolving buried potential landscapes on the scale of tens of nanometers.}

\section{Magnetic Field Discussion}
A weak perpendicular magnetic field $B$ curves the electron trajectories, imparting a shift in transverse momentum $k_{eff} = k_y - \Delta k_y$ where $\Delta k_y = eBL/\hbar$ over the cavity length, which effectively modifies the incidence angle at the p-n interfaces. For a locally linear potential barrier, the transmission probability $T$ is governed by the transverse wavevector $k_y$:
\begin{equation}
T_{\pm}(k_{eff}) = \exp\left(-\frac{\pi\hbar v_F}{eE}k_{eff}^2\right), \quad R_{\pm} = \text{sgn}~(k_{eff}) \sqrt{1-T^2} \label{T}
\end{equation} 
where $E$ is the electric field across the junction \cite{sCheianov_2006}. This relation encapsulates the chiral nature of graphene carriers: trajectories near normal incidence transmit with unit probability (Klein tunneling), while oblique modes are reflected, leading to strong collimation~\cite{sCheianov_2006, sKatsnelson_2006, sShytov_2008, sZhang_2008, sBeenakker_2008, sYoung_2009, sAllain_2011}. The amplitude of the conductance oscillations scales with the interference visibility factor $|T_+|^2|T_-|^2 R_+ R_-$, where the subscripts denote the two cavity interfaces~\cite{sYoung_2009}. At zero magnetic field, the dominant normal modes are highly transmissive ($R \approx 0$), which paradoxically suppresses the interference amplitude. A weak magnetic field shifts the momentum distribution, allowing carriers to access a regime of finite reflection while maintaining significant transmission. This optimization of the product $|T_+|^2|T_-|^2 R_+ R_-$ leads to the observed enhancement of fringe visibility at low fields. However, as the field increases, the cyclotron radius $r_c = \hbar k_F / (eB)$ becomes comparable to the cavity length $L$. in this regime, the trajectories are bent sufficiently that carriers fail to reach the opposing interface, causing the Fabry-Pérot oscillations to extinguish.

\rev{Analogous to monolayer graphene, the decoupled layers of large-angle twisted bilayer graphene are expected to exhibit a $\pi$-phase shift driven by Klein tunneling. The critical magnetic field required for the onset of this shift is inversely proportional to the cavity length ($B \propto 1/L$). As shown in Figure 3a, this phase signature manifests experimentally at approximately 50 mT for the longer cavity mode (circle marker, $L \approx 500$ nm) and at roughly 70 mT for the shorter cavity mode (triangle marker, $L \approx 350$ nm). These values precisely match the expected geometric scaling ratio ($50 \text{ mT} \times 500 / 350 \approx 71$ mT) and are fully corroborated by the electrostatic simulations presented in Figure 3b. While a similar $\pi$-phase shift is distinctly resolved in Figure 3c and d for the circle and triangle regions, this signature is washed out in the square region; the simultaneous presence of three distinct cavity modes in this regime results in a dense superposition that effectively masks the individual phase shifts.}

\end{document}